\documentclass[11pt]{elsarticle}
\usepackage{graphicx} 
\usepackage{natbib}
\usepackage{xcolor}
\usepackage[margin=2.5cm]{geometry}
\usepackage{amsmath}
\usepackage{amsfonts}
\usepackage{amssymb}

\usepackage{hyperref}
\hypersetup{
	colorlinks,
	linkcolor={blue!50!black},
	citecolor={blue!50!black},
	urlcolor={blue!80!black},
	anchorcolor = {blue!80!black},
	filecolor = {blue!80!black},
	menucolor = {blue!80!black},
	runcolor = {blue!80!black}
}
\usepackage{cleveref}
\makeatletter
\def\ps@pprintTitle{%
  \let\@oddhead\@empty
  \let\@evenhead\@empty
  \def\@oddfoot{\reset@font\hfil\thepage\hfil}
  \let\@evenfoot\@oddfoot
}
\makeatother

\title{A mass-conserving contact line treatment for second-order conservative phase field methods based on the generalized Navier boundary condition}
\author[StanfordMechanicalEngAddress]{Reed L. Brown}
\ead{rbrown97@stanford.edu}
\author[StanfordMechanicalEngAddress]{Shahab Mirjalili}
\ead{ssmirjal@stanford.edu}
\author[StanfordMechanicalEngAddress]{Makrand A. Khanwale}
\ead{khanwale@stanford.edu}
\author[StanfordMechanicalEngAddress]{Ali Mani}
\ead{alimani@stanford.edu}

\address[StanfordMechanicalEngAddress]{Department of Mechanical Engineering, Stanford University, CA, USA 94305}

\begin{document}
\begin{frontmatter}
\begin{abstract}
    A mass-conserving contact line treatment for second-order conservative phase field methods is presented and applied to the conservative diffuse interface (CDI) model. The treatment centers on a no-flux boundary condition for the phase field along with a slip boundary condition for the velocity that is based on the generalized Navier boundary condition (GNBC). Since the CDI model is a second-order partial differential equation, it does not permit a second (contact angle) boundary condition, in contrast to the popular fourth-order Cahn-Hilliard model. As such, we use one-sided stencils and extrapolations from the interior of the domain to compute phase-field-related quantities on and near the wall. Additionally, we propose novel modifications to the GNBC on the continuous and discrete levels that reduce spurious slip velocity when the contact angle achieves its equilibrium value. The proposed treatment is validated with the equilibrium drop and two-phase Couette flow test cases.
\end{abstract}
\end{frontmatter}

\section{Introduction}
In the field of two-phase flow modeling, a longstanding focus has been the treatment of the contact line, which is defined as the intersection of a fluid-fluid interface with a solid wall boundary. Contact lines play an important role in the fluid mechanics of a variety of applications in nature and industry, including droplet impact, superhydrophobic surfaces, flow in porous media, microfluidics, and inkjet printing, to name a few \citep{snoeijer2013moving, sui2014numerical,mohammad2022review}. As a result, various treatments have been proposed, in the context of various two-phase flow methods, that attempt to accurately model the physics of contact lines. For a comprehensive summary of these treatments, the reader is referred to the review articles of \citep{snoeijer2013moving, sui2014numerical,mohammad2022review}. A central point to all of these treatments is that the traditional no-slip boundary condition leads to a discontinuous velocity and a shear stress singularity at the contact line \citep{huh1971hydrodynamic, dussan1976, dussan1979spreading}. Therefore, a key requirement for contact line models is that they introduce some mechanism for regularization, and a historically popular choice has been to use a slip boundary condition (see e.g. \citep{hocking1977moving,huh1977steady,zhou1990dynamics}). In addition, the contact angle, which is the angle between the fluid-fluid interface and the solid wall, plays a crucial role in the flow around contact lines. It is important to distinguish between the equilibrium or static contact angle $\theta_{eq}$ and the actual or dynamic contact angle $\theta$. The equilibrium contact angle is a material property, which can be calculated using Young's relation (based on the balance of surface tension forces) \citep{mohammad2022review}. In a static system, the actual contact angle achieves the equilibrium value, but in the presence of flow, the actual contact angle deviates from the equilibrium value. Therefore, a second important requirement for contact line treatments is that they accurately model both static and dynamic contact angles. A common model, which arises from several different physical arguments, is that the contact line speed is proportional to $\cos(\theta)-\cos(\theta_{eq})$ \citep{sui2014numerical}. Lastly, mass conservation is also required. A two-phase flow method that exhibits mass conservation without contact lines should not lose that property after being augmented with a contact line model.

A prerequisite for numerical simulations of contact lines is a framework for the simulation of two-phase flows, for which there exist a variety of techniques. We briefly summarize them here. Broadly speaking, there are two main branches: interface tracking methods and interface capturing methods \citep{Mirjalili_ARB}. The former involves explicit tracking of the location of the interface, and examples include the front-tracking and marker-and-cell methods \citep{mckee2008mac,tryggvason2011direct}. The latter forgoes explicit tracking and instead represents the interface through a mathematical field that is evolved according to a partial differential equation (PDE) \citep{prosperetti2009computational}. Interface capturing methods are further subdivided into two classes: 1) sharp interface methods such as the volume-of-fluid and level set methods (see \citep{prosperetti2009computational,scardovelli2003interface,pilliod2004second,zhang2014refined,olsson2007conservative,osher2004level} and references therein), and 2) diffuse interface or phase field methods (see \citep{chiu2011conservative,mirjalili2020conservative,ten2023unified,shen2015decoupled,khanwale2022fully,khanwale2023projection,huang2020consistent} and references therein).

In this work, we focus on the phase field approach, which uses a smooth ``phase field" to represent the interface as a computationally tractable diffuse region. The phase field evolves according to a PDE that includes an advective term as well as a regularization term that maintains the thickness of the diffuse interface to be on the order of a prescribed length scale parameter. This parameter is chosen to be large enough such that the interface is resolved by the mesh. Away from the interface, the phase field saturates to a constant value, indicating pure fluid 1 or pure fluid 2. The phase field PDE is coupled with equations for continuity and the mixture momentum to simulate two-phase flows. Several different phase field PDEs have been used in the literature, and examples include the Cahn-Hilliard \citep{ten2023unified,shen2015decoupled,khanwale2023projection}, Allen-Cahn \citep{allen1979microscopic}, Navier-Stokes-Korteweg \citep{gomez2023phase}, and second-order conservative phase field \citep{chiu2011conservative,mirjalili2021consistent,huang2020consistent} models. 

Recently, second-order conservative phase field methods have increased in popularity. Based on the Allen-Cahn model but modified to ensure mass conservation (in the absence of source or sink terms), these methods employ a second-order spatial operator and are attractive due to their provable boundedness of the phase field, flexibility in coupling with other multiphysics models, and lower computational complexity compared to the fourth-order Cahn-Hilliard (CH) model, which leads to more efficient and scalable numerical implementations \citep{chiu2011conservative,mirjalili2021consistent,huang2020consistent,rossicomparative}. In this work, we focus on a specific second-order conservative phase field method known as the conservative diffuse interface (CDI) model. CDI has been successfully used to model two-phase flows in both incompressible~\citep{mirjalili2021consistent} and compressible regimes~\citep{jain2020conservative,collis2024diffuse}, and with multiphysics coupling such as scalar transport~\citep{mirjalili2022computational}. However, due to the recent emergence of CDI, a contact line treatment that conserves mass while accurately modeling static and moving contact lines is lacking. In this work, we propose such a contact line treatment. 

Significant work on contact line modeling has been done for the CH model. Because the CH model employs a fourth-order spatial operator, two boundary conditions on the phase field variable may be applied on solid walls. In order to maintain mass conservation, one boundary condition enforces zero regularization flux through the wall. The second boundary condition is then used to enforce a prescribed contact angle or a model based on wall energy relaxation. In addition, the diffuse nature of the interface itself regularizes the singularity, so a no-slip boundary condition is used for the mixture momentum \citep{seppecher1996moving,jacqmin2000contact,chen2000interface, yue2010sharp,yue2011wall}.

A key difference between CH and CDI is that CDI employs a second-order spatial operator, meaning only one boundary condition can be enforced at wall boundaries. As a result, it is not immediately clear how to model contact lines while maintaining mass conservation. One solution is described in the recent work by \citep{huang2022implementing}, who enforce a contact angle boundary condition and compensate for the resulting mass change by using a Lagrange multiplier. However, the challenge with this strategy is that the enforced mass conservation is not necessarily local, since the compensation is applied throughout the domain (a modification that constrains mass conservation to each phase and the wall was proposed in \citep{scapin2022mass}). Furthermore, the contact angle boundary condition used in \citep{huang2022implementing} prescribes the equilibrium contact angle and does not allow deviation of the contact angle from its equilibrium value. Similar work on contact angle boundary conditions for second-order conservative phase field methods was done in \citep{shen2024comparison}.

An alternative strategy is to incorporate contact line physics through a slip boundary condition on the mixture velocity, so that the single boundary condition for CDI can be used to enforce mass conservation. In this work, we use a slip boundary condition based on the generalized Navier boundary condition (GNBC), which was first introduced in \citep{qian2003molecular,qian2005molecular}. From careful analysis of molecular dynamics data of steady moving contact lines, \citep{qian2003molecular} observed near-complete slip at the contact line (consistent with previous studies \citep{koplik1988molecular,koplik1989molecular,thompson1989simulations,thompson1993microscopic}) and showed that the GNBC accurately models the observed slip. They then implemented the GNBC in the context of the CH model \citep{qian2003molecular}, and later showed that for the CH model, the GNBC can also be derived using variational arguments \citep{qian2006variational}. Additionally, the GNBC has since been used more broadly in a variety of other two-phase flow methods \citep{gerbeau2009generalized,yamamoto2013numerical,smuda2021extended}. The GNBC prescribes a slip velocity that is proportional to the total hydrodynamic stress, which includes the viscous stress (used in the standard Navier slip boundary condition) as well as the so-called uncompensated Young's stress, which represents an imbalanced surface tension force density arising from the deviation of the contact angle from its equilibrium value. In this framework, the GNBC for the mixture velocity is used to model contact line physics, in contrast to other treatments that use a contact angle boundary condition for the phase field. A GNBC-based contact line treatment is therefore ideally suited for second-order conservative phase field methods for which only one (mass conservation) boundary condition may be enforced.
    
 In this work, we apply the GNBC to the CDI model of \citep{chiu2011conservative,mirjalili2021consistent}. An early version of this is described in \citep{browngeneralized}. In contrast to the contact angle boundary condition and Lagrange multiplier treatment of \citep{huang2022implementing}, our proposed treatment uses a no-flux boundary condition on the phase field to enforce mass conservation and the GNBC on velocity to model contact line physics. The GNBC allows the contact angle to deviate from its equilibrium value and results in contact line motion that incorporates a physically accurate relationship between contact line speed and contact angle. Indeed, the GNBC contains the $\cos(\theta)-\cos(\theta_{eq})$ functionality. In terms of discretization, the lack of a contact angle boundary condition means that ghost values of the phase field are not computed, and we therefore use one-sided stencils and extrapolation from the interior of the domain to compute phase-field-related quantities on and near the wall. Furthermore, we show that the original version of the GNBC exhibits spurious slip velocity for an equilibrium drop, and we propose modifications to mitigate this issue. Lastly, we demonstrate convergence of our contact line treatment as both the mesh size and interface thickness are refined.

The paper is organized as follows. The two-phase flow method consisting of the CDI model and the Navier-Stokes equations with surface tension, along with the no-flux and slip boundary conditions, are introduced in section \ref{sec:proposed_model}. The slip boundary condition is explained in detail in section \ref{sec: slip boundary condition}. The numerical discretization is described in section \ref{sec: discretization}. Results for two common contact line test cases (equilibrium drop and two-phase Couette flow) are provided in section \ref{sec:results}. Finally, conclusions are offered in section \ref{sec:conclusions}.

\section{Proposed model}
\label{sec:proposed_model}

We use the conservative, second-order-in-space PDE commonly referred to as the Conservative Diffuse Interface (CDI) or Conservative Allen-Cahn model \citep{chiu2011conservative, mirjalili2020conservative} for two-phase flows,

\begin{equation}
    \frac{\partial \phi }{\partial t} + \nabla \cdot (\vec{u}\phi) = \gamma \nabla \cdot \left[ \epsilon \nabla\phi - \phi(1-\phi) \vec{n}_\phi \right]=\nabla\cdot\vec{R},
    \label{eq: CDI}
\end{equation}
where the phase field variable $\phi$ in this case represents the volume fraction of fluid 1 and varies between $0$ and $1$, with $\phi=1$ representing pure fluid $1$ and $\phi=0$ representing pure fluid $2$. Additionally, $\gamma$ is a spatially uniform parameter that controls the strength of the regularization term, $\epsilon$ is a parameter that controls the equilibrium interface thickness, and  $\vec{n}_\phi = \nabla\phi/|\nabla\phi|$ is the interface normal vector. The regularization flux is $\vec{R}= \gamma \left[ \epsilon \nabla \phi - \phi(1-\phi)\vec{n}_\phi \right]$. 

Following \citep{mirjalili2021consistent}, Equation \ref{eq: CDI} is coupled with the incompressible Navier-Stokes equations for the mixture momentum,
\begin{equation}
    \frac{\partial(\rho\vec{u})}{\partial t} + \nabla \cdot \left[ (\rho\vec{u} - \vec{S}) \otimes \vec{u} \right] = -\nabla P + \nabla \cdot [\mu(\nabla\vec{u} + \nabla^T\vec{u})] + \vec{F}_{ST},
    \label{eq: NS_momentum}
\end{equation}
\begin{equation}
    \nabla\cdot \vec{u} = 0.
    \label{eq: NS_incompressibility}
\end{equation}
The mixture density and viscosity are given by $\rho=(\rho_1-\rho_2)\phi + \rho_2$ and $\mu=(\mu_1-\mu_2)\phi + \mu_2$, where subscript $1$ denotes fluid $1$ properties and subscript $2$ denotes fluid $2$ properties. The regularization mass flux, $\vec{S}=(\rho_1-\rho_2)\vec{R}$, is included in the momentum advection term to consistently account for momentum transfer due to the regularization term.

The energy-based (EB) surface tension model is used, following \citep{mirjalili2023assessment},
\begin{equation}
    \vec{F}_{ST}=\mu\nabla\phi,
    \label{eq: EB surface tension}
\end{equation}
where $\mu$ is the chemical potential. In this work, instead of using the well-known form of the chemical potential for the Ginzburg-Landau free energy \citep{mirjalili2023assessment,gurtin1996generalized, abels2012thermodynamically}, given by $$
\mu=6\sigma \left[ \frac{\phi(1-\phi)(1-2\phi)}{\epsilon} - \epsilon\nabla^2\phi \right],
$$ we use a novel, analytically equivalent form given by 
\begin{equation}
    \mu=6\sigma \left[ \frac{\phi(1-\phi)(1-2\phi)(1-|\nabla\psi|^2)}{\epsilon} - \phi(1-\phi)\nabla^2\psi \right],
    \label{eq: chemical potential}
\end{equation}
where $\psi=\epsilon \ln(\phi/(1-\phi))$ is the approximate signed distance function \citep{chiodi2017reformulation,shukla2014nonlinear,waclawczyk2015consistent,jain2022accurate}. This form is be obtained by computing 
\begin{equation}
\nabla\phi=\frac{\phi(1-\phi)}{\epsilon}\nabla\psi
\label{eq: grad_phi in terms of grad_psi}
\end{equation}
and
\begin{equation}
\nabla^2\phi=\nabla\cdot\nabla\phi=\nabla\cdot\left[ \frac{\phi(1-\phi)}{\epsilon}\nabla\psi \right] = \frac{\phi(1-\phi)}{\epsilon}\nabla^2\psi + \frac{\phi(1-\phi)(1-2\phi)}{\epsilon^2}|\nabla\psi|^2
    \label{eq: laplacian_phi in terms of laplacian_psi}
\end{equation}
We have found this form to be discretely advantageous due to the higher accuracy in numerically computing derivatives on the smoother field of $\psi$ rather than $\phi$, especially near the contact lines. In Equation \ref{eq: chemical potential}, $\sigma$ is the surface tension coefficient between fluid $1$ and fluid $2$.

The boundary condition for the phase field at wall boundaries is a no-flux boundary condition,
\begin{equation}
    \vec{R} \cdot \vec{n}_{wall} = 0,
    \label{eq:no-flux BC}
\end{equation}
where $\vec{n}_{wall}$ is the inward wall-normal unit vector. This boundary condition sets the wall-normal regularization flux to be zero. 

The boundary condition for the wall-normal velocity at wall boundaries is the no-penetration condition,
\begin{equation}
    \vec{u} \cdot \vec{n}_{wall} = 0.
    \label{eq:no-penetration BC}
\end{equation}

Equations \ref{eq:no-flux BC} and \ref{eq:no-penetration BC} ensure that there is no flux of $\phi$ into or out of the domain through wall boundaries. Along with appropriate boundary conditions for other domain boundaries (periodic, for example), the total amount of $\phi$ in the domain is conserved. By extension, the total amount of mass in the domain is also conserved.

The boundary condition for the wall-tangential velocity at wall boundaries is a localized slip boundary condition closely related to the generalized Navier boundary condition \citep{qian2003molecular},

\begin{equation}
    u_{slip} = u_{slip}^v + u_{slip}^Y,
    \label{eq: slip BC}
\end{equation}
where
\begin{equation}
    u_{slip}^v = \frac{\mu}{\beta}\frac{\partial u_\tau}{\partial n_{wall}}
    \label{eq: u_slip_v}
\end{equation}
and
\begin{equation}
    u_{slip}^Y = -\frac{6\sigma}{\beta} \left[ \phi(1-\phi)\frac{\partial\psi}{\partial n_{wall}} + \phi(1-\phi)(1+\kappa_p\psi) \cos(\theta_{eq}) \right] \frac{\phi(1-\phi)}{\epsilon} \frac{\partial\psi}{\partial \tau_{wall}}.
    \label{eq: u_slip_Y}
\end{equation}
The slip velocity is defined as $u_{slip}=u_\tau-U_{wall}$, where $u_\tau=\vec{u}\cdot\vec{\tau}_{wall}$ is the wall-tangential component of velocity, $\vec{\tau}_{wall}$ is the wall-tangential unit vector, and $U_{wall}$ is the prescribed wall-tangential velocity of the wall boundary. $\beta$ is the wall-fluid friction parameter and is in general a function of $\phi$. Similar to density and viscosity, we use the linear function given by $\beta=(\beta_1-\beta_2)\phi+\beta_2$ in this work. $\theta_{eq}$ is the equilibrium contact angle, which is assumed to be a known parameter in this work, determined by the material properties of the fluids and wall. $\kappa_p$ is the interface curvature in the plane containing $\vec{n}_\phi$ and $\vec{n}_{wall}$. For the remainder of the paper, we will restrict ourselves to a two-dimensional setting, where $\vec{n}_\phi$ and $\vec{n}_{wall}$ are in the $x-y$ plane, and $\kappa_p$ is equal to the full interface curvature $\kappa$, given by

\begin{equation}
    \kappa=\frac{1-2\phi}{\epsilon} \left[ 1-|\nabla\psi|^2 \right] - \nabla^2\psi,
    \label{eq:curvature}
\end{equation}
which is the curvature implied by the EB model, given in Equations \ref{eq: EB surface tension} and \ref{eq: chemical potential}. This curvature estimate results from the equivalency of the EB model with the localized CSF model at phase field equilibrium, as shown in \cite{mirjalili2023assessment}.

The viscous slip $u_{slip}^v$ models slip that is proportional to the viscous stress and corresponds to the Navier slip boundary condition. The contact line or Young's slip $u_{slip}^Y$ models slip that is proportional to the uncompensated Young's stress, which arises from the deviation of the contact angle from its equilibrium value. In addition, $u_{slip}^Y$ incorporates novel modifications to the GNBC that reduce spurious slip at equilibrium. A more detailed explanation is provided in section \ref{sec: slip boundary condition}.

\section{Slip boundary condition}
\label{sec: slip boundary condition}

The proposed slip boundary condition in this work is an adaptation of the generalized Navier boundary condition (GNBC) introduced by \cite{qian2003molecular}. The authors' central finding, which was informed by analysis of molecular dynamics simulations, is that the slip velocity $u_{slip}$ near a contact line is proportional to the total hydrodynamic stress, which is the sum of the viscous stress $\sigma^v$ (corresponding to the Navier slip boundary condition) and the so-called uncompensated Young's stress $\tilde{\sigma}^Y$, which arises from the deviation of the actual contact angle from its equilibrium value, 
\begin{equation}
    \beta u_{slip} = \sigma^v + \tilde{\sigma}^Y
    \label{eq: GNBC_viscous_plus_uncompensated_Young's},
\end{equation}
where $\beta$ is a proportionality constant that represents wall-fluid friction. This can be equivalently written, as in Equation \ref{eq: slip BC},
$$
    u_{slip} = u_{slip}^v + u_{slip}^Y,
$$
where the viscous slip is $u_{slip}^v=\sigma^v/\beta$ and the contact line or Young's slip is $u_{slip}^Y=\tilde{\sigma}^Y/\beta$.

The viscous stress is given by $\sigma^v=\mu(\nabla\vec{u} + (\nabla\vec{u})^T)$.  For this work, we consider no-penetration wall boundaries, so the viscous stress reduces to 
\begin{equation}
    \sigma^v=\mu \frac{\partial u_\tau}{\partial n_{wall}}.
    \label{eq: sigma_v}
\end{equation} 

\begin{figure}
    \centering
    \includegraphics[width=.7\linewidth]{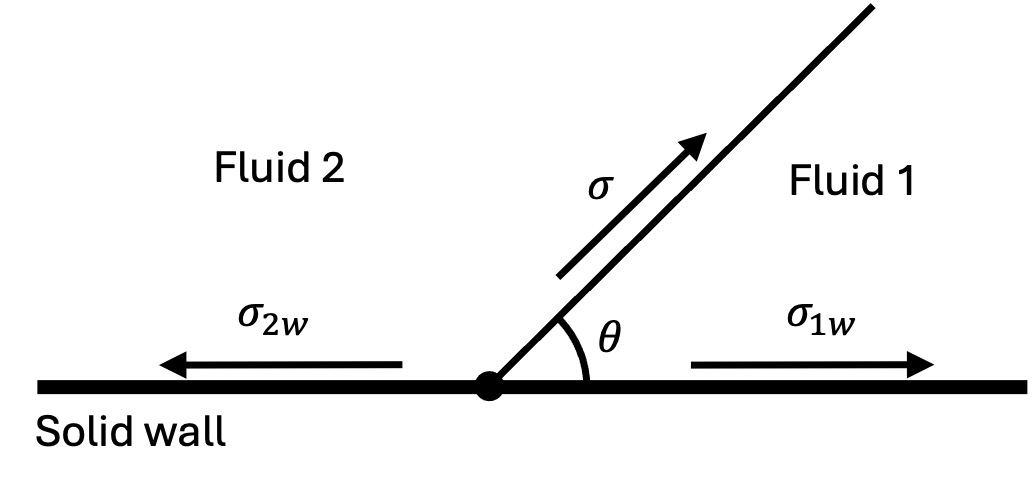}
    \caption{Schematic of the surface tension forces acting on a contact line, in a sharp interface representation. The two wall-fluid surface tensions act opposite each other in the wall-tangential direction. The fluid-fluid surface tension has a $\cos(\theta)$ component in the wall-tangential direction, where $\theta$ is the contact angle. The net wall-tangential force is $\Sigma F^{ST}=\sigma\cos(\theta)+\sigma_{1w}-\sigma_{2w}$.}
    \label{fig: sharp interface contact line}
\end{figure}

The uncompensated Young's stress stems from surface tension imbalance and can be intuitively understood by first considering the surface tension forces acting on a sharp interface contact line. A schematic of this is shown in Figure \ref{fig: sharp interface contact line}. There exist surface tension forces due to the interfaces between fluid 1 and fluid 2 ($\sigma_{12}=\sigma$), fluid 1 and the solid wall ($\sigma_{1w}$), and fluid 2 and the solid wall ($\sigma_{2w}$). If the fluid-fluid interface intersects the solid wall at a contact angle $\theta$, then the sum of wall-tangential forces acting on the contact line is $\Sigma F^{ST}=\sigma\cos(\theta) + \sigma_{1w} - \sigma_{2w}$. Young's relation considers the equilibrium situation in which $\Sigma F^{ST}=0$ and the contact angle achieves its equilibrium value $\theta_{eq}$,
\begin{equation}
    \sigma\cos(\theta_{eq}) = -(\sigma_{1w}-\sigma_{2w}) = -\Delta\sigma_w.
    \label{eq: Young's relation}
\end{equation}
Therefore, the net force on the contact line can be written
\begin{equation}
    \Sigma F^{ST} = \sigma\cos(\theta) - \sigma\cos(\theta_{eq}).
    \label{eq: net ST force}
\end{equation}

The connection between this net force and the uncompensated Young's stress is that the integral of the latter across the contact line region must equal the former,
\begin{equation}
    \int_{interface}\tilde{\sigma}^Y d\tau_{wall} = \Sigma F^{ST} = \sigma (\cos(\theta)-\cos(\theta_{eq})).
    \label{eq: integral constraint sigma}
\end{equation}

Or in terms of $u_{slip}^Y$,
\begin{equation}
    \int_{interface}u_{slip}^Y d\tau_{wall} = \frac{\sigma}{\beta} (\cos(\theta)-\cos(\theta_{eq})).
    \label{eq: integral constraint u_slip}
\end{equation}

The findings of \cite{qian2003molecular} therefore state that, in the presence of contact lines, the true slip velocity is composed of the usual viscous slip $u_{slip}^v$ as well as a contact line slip $u_{slip}^Y$, whose integral across the contact line region is proportional to the surface tension imbalance arising from the deviation of the actual contact angle from its equilibrium value. A similar interpretation of the GNBC is provided in \cite{qian2003molecular}, but we restate it here to emphasize this important point.

In the remainder of this section, we will introduce several potential options for $u_{slip}^Y$, which differ in certain aspects but which all satisfy Equation \ref{eq: integral constraint u_slip}. To derive these options, it is useful to start with a general form for $u_{slip}^Y$,
\begin{equation}
    u_{slip}^Y = f(\phi)[\cos(\theta)-\cos(\theta_{eq})],
    \label{eq: uncompensated Young's general form}
\end{equation}
where $f(\phi)$ is a localized function that satisfies $\int_{interface}f(\phi)d\tau_{wall}=\sigma/\beta$. The first step in obtaining a suitable boundary condition for computation is to write $\cos(\theta)$ in terms of $\phi$, which can be achieved by noting that at all points on the wall, $-\partial\phi/\partial n_{wall} = |\nabla\phi|\cos(\theta) $, so 
$$
u_{slip}^Y=-f(\phi)\left[ \frac{\partial\phi/\partial n_{wall}}{|\nabla\phi|} +\cos(\theta_{eq})\right].
$$
Next, if one chooses $f(\phi)=(6\sigma/\beta)\epsilon|\nabla\phi| \partial\phi/\partial \tau_{wall}$, then

\begin{equation}
\text{\textbf{Model 1B}: }\quad u_{slip}^Y=-\frac{6\sigma}{\beta}\left[ \epsilon\frac{\partial\phi}{\partial n_{wall}} +\epsilon|\nabla\phi|\cos(\theta_{eq})\right] \frac{\partial \phi}{\partial \tau_{wall}},
\label{eq: Model 1B}
\end{equation}
which we denote as Model 1B.

A slightly different model arises from assuming that the interface is at phase field equilibrium, such that $\epsilon|\nabla\phi|=\phi(1-\phi)$. This leads to a slip model we denote as Model 1A,
\begin{equation}
\text{\textbf{Model 1A}: }u_{slip}^Y = - \frac{6\sigma}{\beta} \left[ \epsilon \frac{\partial\phi}{\partial n_{wall}} + \phi(1-\phi)\cos(\theta_{eq}) \right] \frac{\partial\phi}{\partial\tau_{wall}}.   
    \label{eq: Model 1A}
\end{equation}

Model 1A was used in our previous work \cite{browngeneralized} and is very similar to the model proposed in \cite{qian2003molecular}. The main difference is due to the different conventions (bounds of $\phi$ and definition of interface thickness parameter) between the Cahn-Hilliard model, which was used in \cite{qian2003molecular}, and the CDI model used in the present work. We found in \cite{browngeneralized} that Model 1A produces good results, but exhibits shortcomings that will be elaborated on in Section \ref{subsec: prescribed phase field}.

\subsection{Analysis of slip models using a prescribed equilibrium phase field}
\label{subsec: prescribed phase field}

Since the contact line slip $u_{slip}^Y$ depends only on the phase field $\phi$, one simple and useful way to assess different slip models is to analytically prescribe $\phi$ for an equilibrium solution ($\theta=\theta_{eq})$ and then compute the resulting $u_{slip}^Y$, which should ideally be zero everywhere. Indeed, using this analysis we show below that Models 1A and 1B exhibit ``spurious" slip, i.e. non-zero slip velocity for the equilibrium solution, and that this spurious slip is due to a combination of discretization errors and model errors.

\begin{figure}
    \centering
    \includegraphics[width=.7\linewidth]{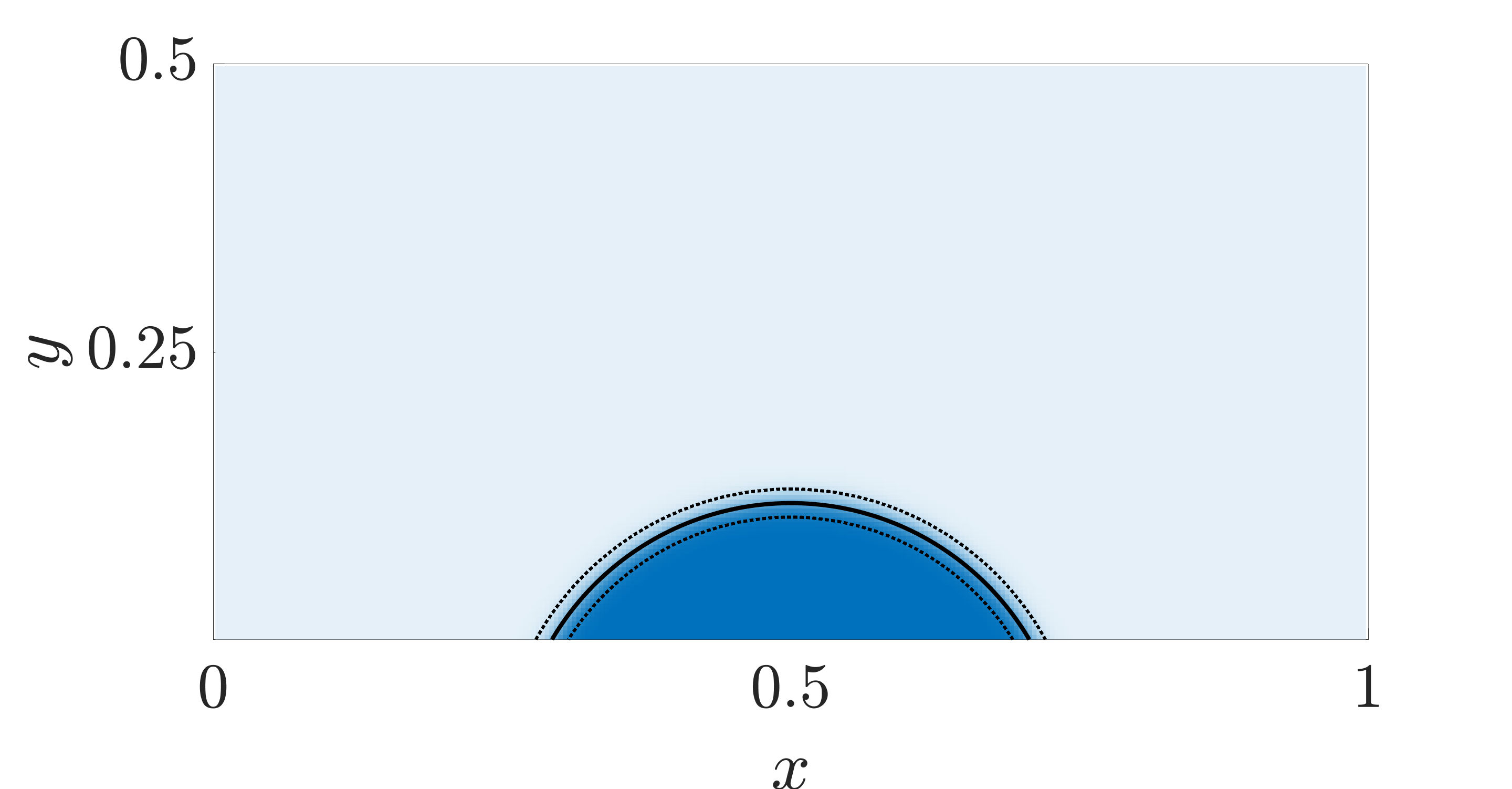}
    \caption{A prescribed equilibrium phase field $\phi(x,y)$ representing a drop that forms a $\theta=60^\circ$ contact angle with the wall. The drop is circular, corresponding to capillary equilibrium, and the interface is at phase field equilibrium. As a result, the contours of the phase field are concentric circular arcs. The thick black line is the $\phi=0.5$ contour, while the inner dashed line is the $\phi=0.9$ contour and the outer dashed line is the $\phi=0.1$ contour.}
    \label{fig: prescribed phase field}
\end{figure}

This analysis is done in a two-dimensional domain $x\in[0,L_x]$ and $y\in[0,L_y]$, with $L_x=1$ and $L_y=0.5$. We prescribe the phase field to be
\begin{equation}
    \phi(x,y) = \frac{1}{2} \left[ 1 + \tanh\left( \frac{R-\sqrt{(x-x_c)^2 + (y-y_c)^2}}{2\epsilon} \right) \right],
    \label{eq: prescribed phase field}
\end{equation}
where $x_c=L_x/2$ and $y_c=-R\cos(\theta)$, and 
$$
R=\sqrt{\frac{A_0}{\theta(\pi/180) - \sin(2\theta)}},
$$
where
$$
A_0 = \frac{\pi R_0^2}{2}
$$
and $R_0=0.15$. $\theta$ is the contact angle, which is defined as the angle that the $\phi=0.5$ contour makes with the wall. A plot of the prescribed phase field is shown in Figure \ref{fig: prescribed phase field}. The formula for $R$ ensures that the area of the prescribed drop is the same as the area of a semi-circular drop ($\theta=90^\circ$) of radius $R_0=0.15$, regardless of the value of $\theta$. The spatial discretization for this section is based on second-order finite differences on a uniform Cartesian staggered grid and is the same as that used in the fully coupled simulations. More details are provided later in Section \ref{sec: discretization}. The interface thickness parameter is chosen in relation to the mesh size, $\Delta$. We show results for $\epsilon=1.4\Delta$, except for the convergence results in Figure \ref{fig: prescribed phase field mesh convergence}, for which $\epsilon$ varies from $0.7\Delta$ to $1.4\Delta$.

\begin{figure}
    \centering
    \includegraphics[width=.8\linewidth]{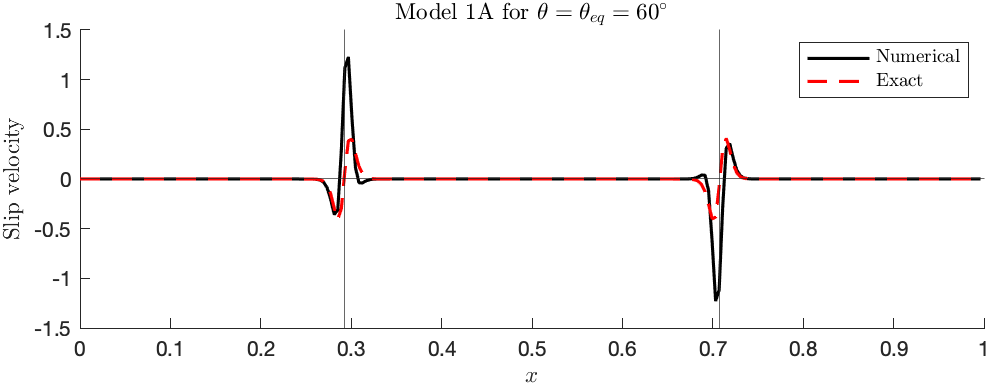}
    \caption{Contact line slip $u_{slip}^Y(x)$, computed using Model 1A, as a function of $x$, for the prescribed equilibrium phase field. The two vertical lines mark the locations where the $\phi=0.5$ contour intersects the wall. There is a significant discrepancy between numerical and exact, and the exact slip is non-zero, indicating both discretization and model error.}
    \label{fig: u_slip(x) for Model 1A}
\end{figure}

In this section, we choose the contact angle to be equal to its equilibrium value, $\theta=\theta_{eq}$. We call this \textit{contact line equilibrium}. Additionally, the circular shape of the drop corresponds to \textit{capillary equilibrium} and the phase field prescribed in Equation \ref{eq: prescribed phase field} results in a regularization flux that is analytically zero, meaning the interface is in \textit{phase field equilibrium}. As a whole, we refer to this as a prescribed equilibrium phase field, where ``equilibrium" takes these three distinct meanings. Given such a phase field, one would expect a good model for $u_{slip}^Y$ to be zero at all points $x$ on the wall. We evaluate this by computing $u_{slip}^Y(x)$ numerically using finite difference approximations, as well as exactly using symbolic math tools (MATLAB Symbolic Math Toolbox). Figure \ref{fig: u_slip(x) for Model 1A} shows the numerical and exact $u_{slip}^Y(x)$ for Model 1A. Two things are clear: First, there is significant discretization error, as evidenced by the large discrepancy between the numerical and exact curves. Second, there is model error, as evidenced by the non-zero exact $u_{slip}^Y(x)$, despite the prescribed equilibrium phase field. We now describe two modifications that reduce the discretization error and eliminate the model error.


\subsubsection{Transforming into an approximate signed distance function reduces discretization error}
The discretization error can be significantly reduced by algebraically transforming $\phi$ into an approximate signed distance function,
\begin{equation}
    \psi=\epsilon \ln \frac{\phi}{1-\phi}.
    \label{eq: psi}
\end{equation}
The gradients of $\phi$ can then be written in terms of the gradient of $\psi$, according to Equation \ref{eq: grad_phi in terms of grad_psi},
$$    
\nabla\phi=\frac{\phi(1-\phi)}{\epsilon} \nabla\psi.
$$
As mentioned previously, the $\psi$ field is smoother than the $\phi$ field, which rapidly varies in the interface region. As a result, finite difference approximations of $\nabla\psi$ exhibit significantly decreased discretization error, compared to approximations of $\nabla\phi$. The transformation from $\phi$ to $\psi$ has been successfully used in previous works \citep{chiodi2017reformulation,shukla2014nonlinear,waclawczyk2015consistent,jain2022accurate}, though to our knowledge, this transformation has never been used in a contact line slip boundary condition.

Model 1A can therefore be transformed using gradients of $\psi$ into the form we refer to as Model 2A,
\begin{equation}
    \text{\textbf{Model 2A}: }\quad u_{slip}^Y = - \frac{6\sigma}{\beta} \left[ \phi(1-\phi) \frac{\partial\psi}{\partial n_{wall}} + \phi(1-\phi)\cos(\theta_{eq}) \right] \frac{\phi(1-\phi)}{\epsilon}\frac{\partial\psi}{\partial\tau_{wall}}.   
\label{eq: Model 2A}
\end{equation}
Similarly, Model 1B is transformed into the form we refer to as Model 2B,
\begin{equation}
    \text{\textbf{Model 2B}: }\quad u_{slip}^Y = - \frac{6\sigma}{\beta} \left[ \phi(1-\phi) \frac{\partial\psi}{\partial n_{wall}} + \phi(1-\phi)|\nabla\psi|\cos(\theta_{eq}) \right] \frac{\phi(1-\phi)}{\epsilon}\frac{\partial\psi}{\partial\tau_{wall}}.
\label{eq: Model 2B}
\end{equation}

\subsubsection{Curvature correction eliminates model error}
Next, we address the model error by introducing a correction to the $\cos(\theta_{eq})$ term in the slip models. We previously defined the contact angle $\theta$ as the angle that the $\phi=0.5$ contour makes with the wall. In this section, we will temporarily use a more general notation that denotes the contact angle for arbitrary contours. Let $\theta_{\phi_0}$ be the angle that the $\phi=\phi_0$ contour makes with the wall. 

\begin{figure}
    \centering
    \includegraphics[width=.7\linewidth]{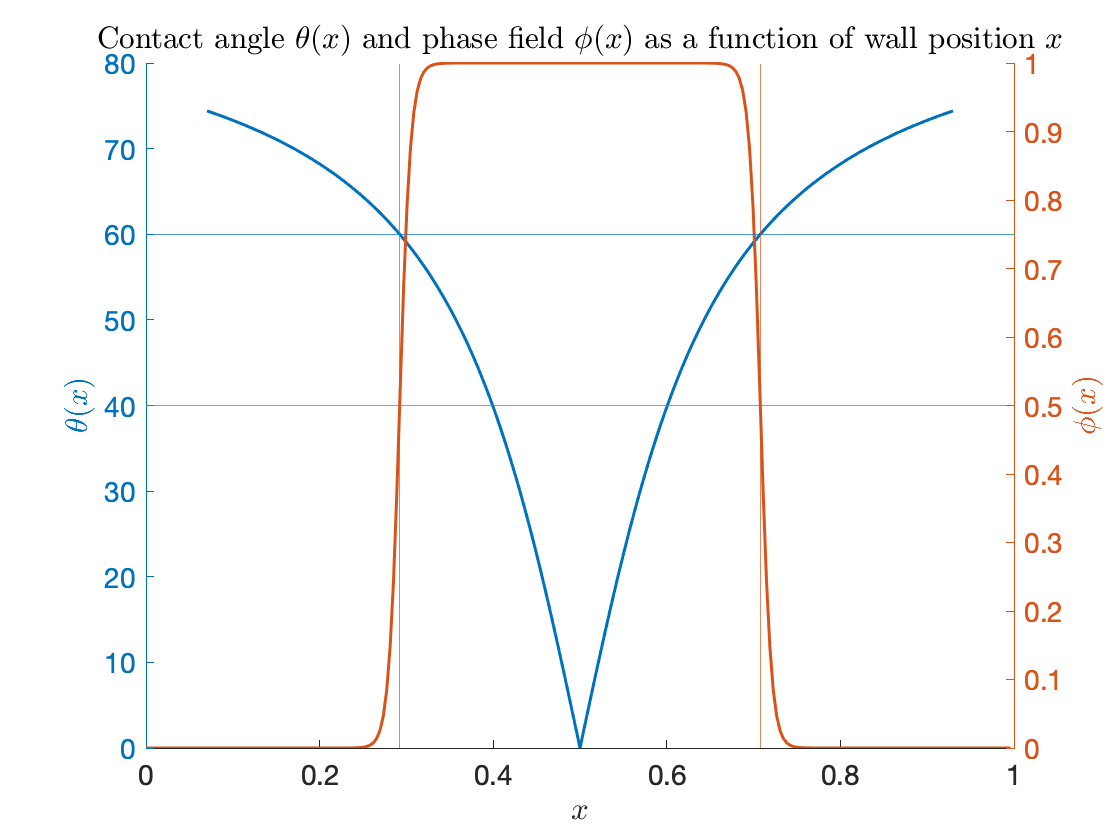}
    \caption{Blue: Contact angle $\theta(x)=\cos^{-1}((-\partial\phi/\partial y) / |\nabla\phi|)$ as a function of wall position $x$, for the prescribed equilibrium phase field. The contact angle is calculated exactly using symbolic math tools. The horizontal line marks $\theta=\theta_{eq}=60^\circ$. Orange: Phase field $\phi(x)$ as a function of wall position $x$. The two vertical lines mark the locations where the $\phi=0.5$ contour intersects the wall. At the locations of the $\phi=0.5$ contour, $\theta=\theta_{eq}=60^\circ$, but at all other locations, the contact angle deviates from $60^\circ$. This variation is a result of the contours' being concentric circular arcs.}
    \label{fig: prescribed phase field theta(x) and phi(x)}
\end{figure}

We remind that for this prescribed equilibrium phase field, all contours of $\phi$ are concentric circular arcs. The $\phi=0.5$ contour has a contact angle $\theta_{0.5}=\theta_{eq}$. However, all other contours necessarily have contact angles different from $\theta_{eq}$. Without loss of generality, assume $\theta_{eq}=60^\circ$. The same argument holds for any other acute contact angle, and an analogous argument holds for obtuse contact angles. If $\theta_{eq}=60^\circ$, the $\phi>0.5$ contours have contact angles less than $60^\circ$ and the $\phi<0.5$ contours have contact angles greater than $60^\circ$, as shown in Figure \ref{fig: prescribed phase field theta(x) and phi(x)}. In this figure, the contact angle is calculated exactly using symbolic math tools as $\theta(x)=\cos^{-1}((-\partial\phi/\partial y)/|\nabla\phi|)$ and is a function of wall position $x$.

\begin{figure}
    \centering
    \includegraphics[width=.7\linewidth]{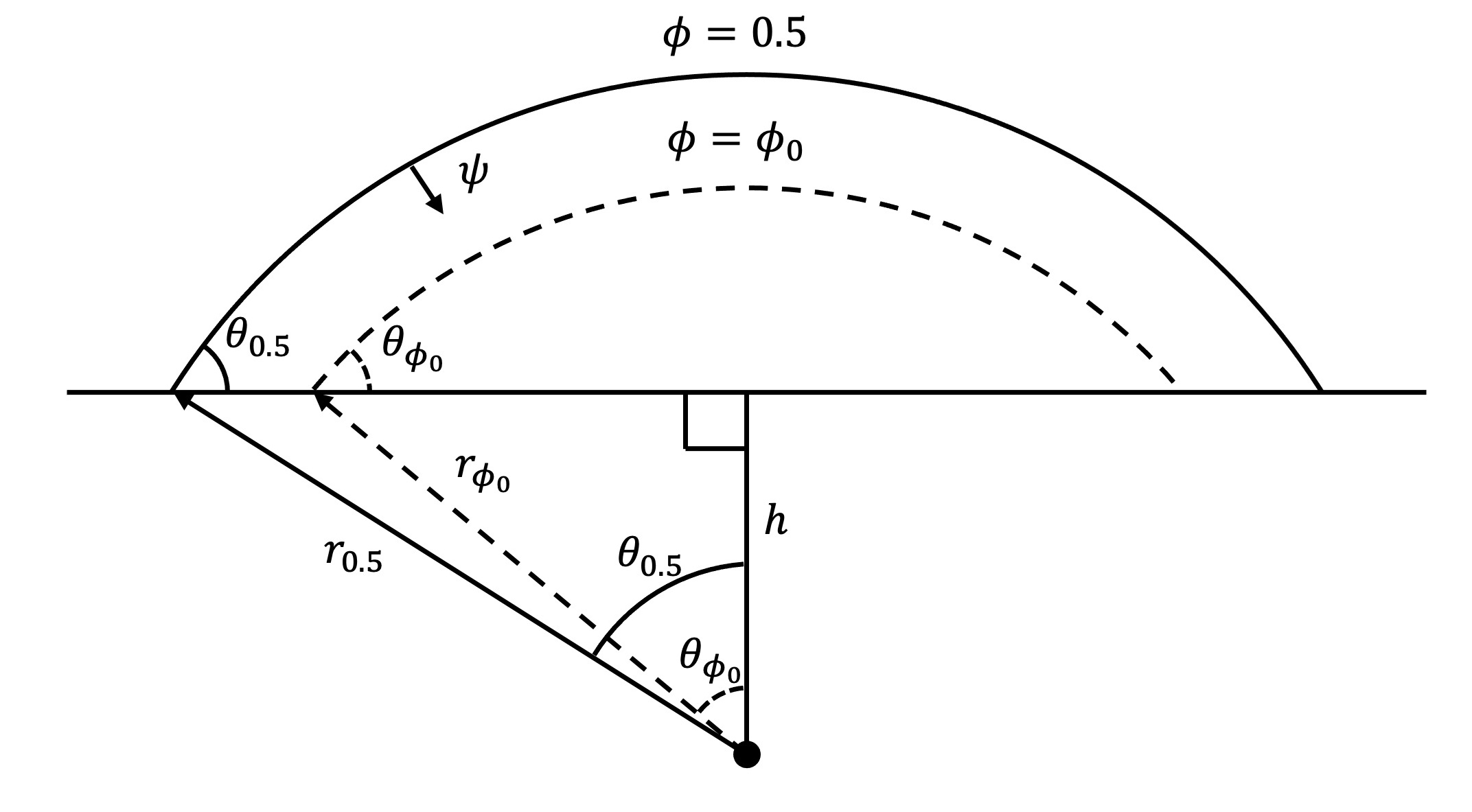}
    \caption{Schematic of how the contact angle of an arbitrary contour $\theta_{\phi_0}$ is related to the contact angle of the $\phi=0.5$ contour $\theta_{0.5}$. By trigonometric relations, one can find that $\cos(\theta_{\phi_0})=(r_{0.5}/r_{\phi_0})\cos(\theta_{0.5})$. This relation informs the curvature correction.}
    \label{fig: curvature correction schematic}
\end{figure}

The value of the contact angle for an arbitrary contour can be determined using the schematic in Figure \ref{fig: curvature correction schematic}. Consider two contours: $\phi=0.5$ and $\phi=\phi_0>0.5$. (One can arrive at the same conclusion by considering $\phi_0<0.5$). The $\phi=0.5$ contour has a radius $r_{0.5}$ and contact angle $\theta_{0.5}=\theta_{eq}$. The $\phi=\phi_0$ contour has a radius $r_{\phi_0}$ and contact angle $\theta_{\phi_0}$. These radii and contact angles are related by 
$$
r_{0.5}\cos(\theta_{0.5})=h=r_{\phi_0}\cos(\theta_{\phi_0}),
$$
so that 
$$
\cos(\theta_{\phi_0})=\frac{r_{0.5}}{r_{\phi_0}}\cos(\theta_{0.5}).
$$ 
The difference between $r_{0.5}$ and $r_{\phi_0}$ is the signed distance, $\psi_{\phi_0}=r_{0.5}-r_{\phi_0}$, so 

\begin{equation}
\cos(\theta_{\phi_0})=\frac{r_{\phi_0}+\psi_{\phi_0}}{r_{\phi_0}}\cos(\theta_{0.5}) = \left( 1+\frac{\psi_{\phi_0}}{r_{\phi_0}} \right) \cos(\theta_{0.5}) = \left(1+\kappa^p_{\phi_0}\psi_{\phi_0} \right) \cos(\theta_{0.5}),
\label{eq: phi_0 contact angle}
\end{equation}
where $\kappa_{p,\phi_0}=1/r_{\phi_0}$ is the curvature of the $\phi_0$ contour in the plane containing $\vec{n}_{\phi}$ and $\vec{n}_{wall}$. Using this information, we introduce a ``curvature correction", $1+\kappa_p\psi$, to transform Models 1A, 1B, 2A, and 2B into Models 3A, 3B, 4A, and 4B, respectively.

\begin{equation}
    \text{\textbf{Model 3A}: }\quad u_{slip}^Y = - \frac{6\sigma}{\beta} \left[ \epsilon \frac{\partial\phi}{\partial n_{wall}} + \phi(1-\phi)(1+\kappa_p\psi)\cos(\theta_{eq}) \right] \frac{\partial\phi}{\partial\tau_{wall}}, 
    \label{eq: Model 3A}
\end{equation}

\begin{equation}
\text{\textbf{Model 3B}: }\quad u_{slip}^Y=-\frac{6\sigma}{\beta}\left[ \epsilon \frac{\partial\phi}{\partial n_{wall}} +\epsilon|\nabla\phi|(1+\kappa_p\psi)\cos(\theta_{eq})\right] \frac{\partial \phi}{\partial \tau_{wall}},
\label{eq: Model 3B}
\end{equation}

\begin{equation}
    \text{\textbf{Model 4A}: }\quad u_{slip}^Y = - \frac{6\sigma}{\beta} \left[ \phi(1-\phi) \frac{\partial\psi}{\partial n_{wall}} + \phi(1-\phi)(1+\kappa_p\psi)\cos(\theta_{eq}) \right] \frac{\phi(1-\phi)}{\epsilon}\frac{\partial\psi}{\partial\tau_{wall}},
\label{eq: Model 4A}
\end{equation}

\begin{equation}
    \text{\textbf{Model 4B}: }\ u_{slip}^Y = - \frac{6\sigma}{\beta} \left[ \phi(1-\phi) \frac{\partial\psi}{\partial n_{wall}} + \phi(1-\phi)|\nabla\psi|(1+\kappa_p\psi)\cos(\theta_{eq}) \right] \frac{\phi(1-\phi)}{\epsilon}\frac{\partial\psi}{\partial\tau_{wall}}.
    \label{eq: Model 4B}
\end{equation}

In this work, we restrict ourselves to a two-dimensional setting in which the in-plane curvature $\kappa_p$ is equal to the full interface curvature $\kappa$, which is given by Equation \ref{eq:curvature}.

The effect of the curvature correction is to adjust the ``target" contact angle for each contour. For the $\phi=0.5$ contour, the target is still $\theta_{eq}$, since $\psi=0$ at the $\phi=0.5$ contour. However, for other contours, the target has been adjusted in order to account for the variation in contact angle inherent to a curved, diffuse interface in phase field equilibrium.

\begin{figure}
    \centering
    \includegraphics[width=.8\linewidth]{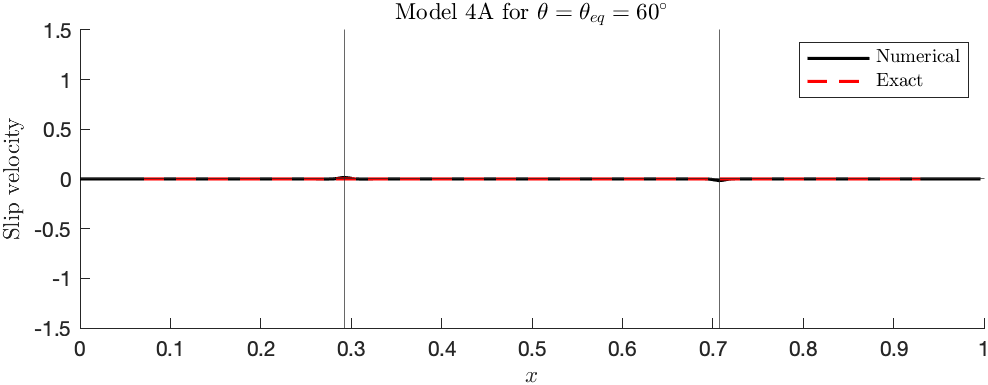}
    \caption{Contact line slip $u_{slip}^Y(x)$, computed using Model 4A, as a function of $x$. The two vertical lines mark the locations where the $\phi=0.5$ contour intersects the wall. Compared to Figure \ref{fig: u_slip(x) for Model 1A}, the exact slip is now zero, and the numerical slip is significantly reduced.}
    \label{fig: u_slip(x) for Model 4A}
\end{figure}

\begin{figure}
    \centering
    \includegraphics[width=.7\linewidth]{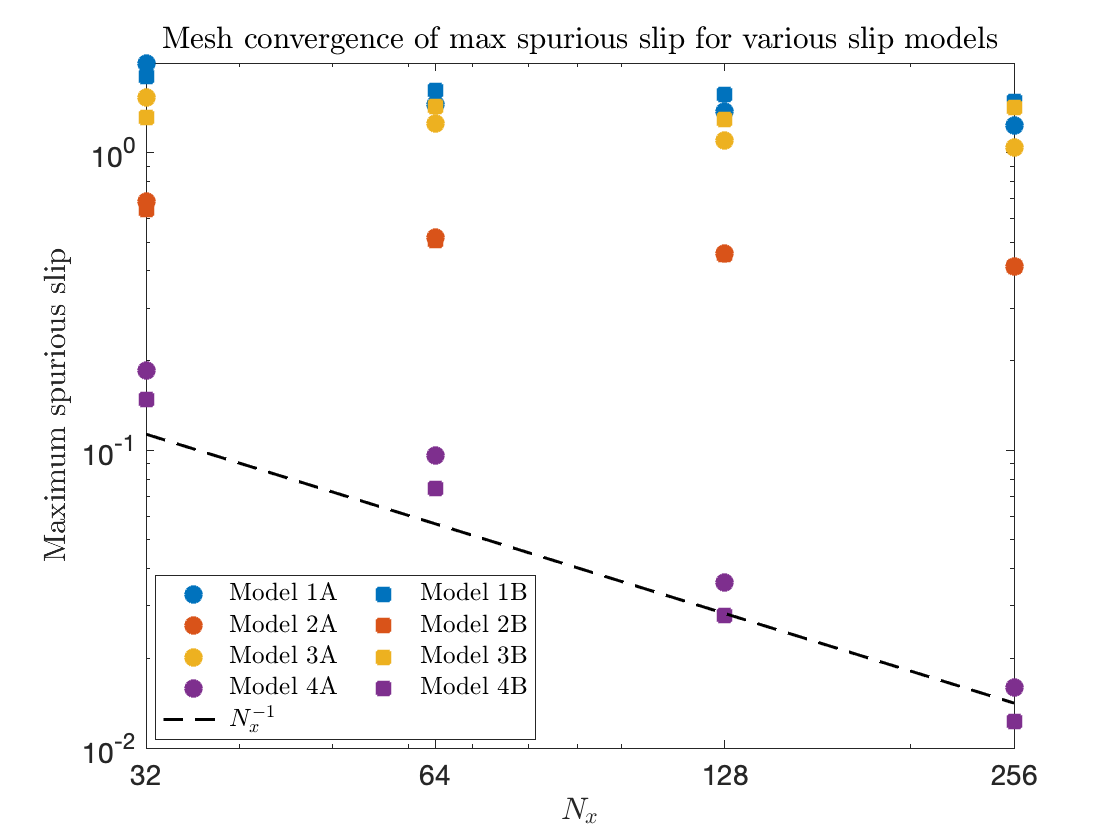}
    \caption{Convergence plot of maximum spurious slip $max(|u_{slip}^Y(x)|)$ as a function of number of points in the $x$ direction $N_x$, for the various slip models presented in this section. In this convergence, the interface thickness parameter $\epsilon$ is decreased along with the mesh size $\Delta$ according to $\epsilon\sim\Delta^{2/3}$. The dashed black line indicates first-order convergence ($N_x^{-1}$).}
    \label{fig: prescribed phase field mesh convergence}
\end{figure}

Figure \ref{fig: u_slip(x) for Model 4A} shows $u_{slip}^Y$ using Model 4A and illustrates the significant decrease in discretization error from using $\psi$, as well as the elimination of model error from the curvature correction. For a more quantitative comparison, Figure \ref{fig: prescribed phase field mesh convergence} shows mesh convergence of the maximum spurious slip $max|u_{slip}^Y|$ for the various models. In this mesh convergence, the interface thickness parameter $\epsilon$ is decreased along with the mesh size $\Delta$, but at a slower rate according to $\epsilon\sim \Delta^{2/3}$. We make this choice in order to simultaneously probe mesh convergence of the discretization as well as approach the sharp interface limit \cite{jacqmin1999calculation,mirjalili2019comparison,mirjalili2023assessment}. 

It is evident that only Models 4A/B show satisfactory convergence. The range of $\epsilon$ values spans from $0.7\Delta$ to $1.4\Delta$, which comprise relatively coarsely resolved interfaces. As a result, discretization errors are large for Models 1A/B and 3A/B. The error is smaller for Models 2A/B, due to reduced discretization error from using $\psi$, but the lack of the curvature correction in these models results in non-convergence. Models 4A/B enjoy reduced discretization error from $\psi$ as well as the curvature correction, resulting in convergence that is slightly faster than first-order. It is worth noting that for larger values of $\epsilon/\Delta$, the discretization error is smaller and the benefit of the curvature correction in Models 3A/B becomes more evident. 

In the remainder of this work, we present results from our fully coupled solver for only Model 4A. This decision is primarily based on the results of this section, which demonstrate that only Models 4A/B exhibit spurious slip convergence for the prescribed equilibrium phase field. The choice of version A over version B is based on the assumption that the interface remains close to phase field equilibrium. This assumption appears to be valid in our simulations. However, with different test cases or parameter values, the interface may deviate more significantly from equilibrium, in which case version B may be superior. On the other hand, the $|\nabla\psi|$ factor brings along some discretization error, which may negate any beneficial effect. Investigation of this, and a comprehensive evaluation of the various models in general, is reserved for future work.

\section{Discretization}
\label{sec: discretization}
For spatial discretization, we use finite differences on a Cartesian staggered grid with uniform mesh spacing $\Delta$. For time integration, we use RK4 time-stepping. More discretization details can be found in \cite{mirjalili2021consistent,mirjalili2024conservative}. Only two-dimensional test cases are presented in this work; three-dimensional test cases are reserved for future work. In the two-dimensional domain, the left and right boundaries are treated as periodic, while the top and bottom boundaries are treated as walls.

Second-order central differences are used in the interior of the domain. Near walls, however, second-order one-sided stencils in the wall-normal direction are used to compute derivatives of $\phi$ or $\psi$, which are required to compute several quantities, including the normal vector $\vec{n}$ and the surface tension body force $\vec{F}_{ST}$. Furthermore, the quantities that appear in the various models for $u_{slip}^Y$ are obtained by first-order linear extrapolation from the interior to the wall. One-sided stencils and extrapolation are used because the second-order CDI model does not permit a second contact angle boundary condition (in addition to the no-flux boundary condition). As a result, the ghost values of $\phi$ are never computed or used, and wall quantities related to the phase field must come from interior data.

\section{Results}
\label{sec:results}
In this section, the proposed contact line treatment using Model 4A for $u_{slip}^Y$ is validated with two test cases, which evaluate distinct features of the model. The equilibrium drop test case \citep{huang2022implementing, shen2024comparison} focuses on how well static contact line physics is captured, whereas the two-phase Couette flow \citep{qian2005molecular,gerbeau2009generalized,smuda2021extended} tests the model's behavior for moving contact lines. The reader is reminded that the results in this section are from a fully-coupled, time-advancing solver, in contrast to the prescribed equilibrium phase field analysis presented in Section \ref{subsec: prescribed phase field}. 

\subsection{Equilibrium drop}
In this test case, a semi-circular drop of radius $R=0.15$ is initialized on the bottom wall in a domain $x\in[0,L_x]$ and $y\in[0,L_y]$, with $L_x=1$ and $L_y=0.5$. The initial velocity field is zero. The top and bottom boundaries are walls, on which the contact line treatment is applied. The left and right boundaries are periodic. The two fluids have identical density $\rho=0.01$ and viscosity $\mu=0.01$, and the surface tension coefficient is $\sigma=1$. The wall-fluid friction parameter is $\beta=10$. The regularization strength is chosen in relation to the maximum velocity magnitude, $\gamma=2.5|\vec{u}|_{max}$, in accordance with \citep{mirjalili2020conservative} to maintain boundedness of $\phi$. All parameter values and results are non-dimensional unless otherwise noted.

The initial phase field is the prescribed equilibrium phase field given in Equation \ref{eq: prescribed phase field} but with an initial contact angle $\theta=\theta_{init}=90^\circ$. The equilibrium contact angle is then set to $\theta_{eq}=60^\circ$, and the system is allowed to evolve to steady state. Since $\theta_{eq}$ is different from $\theta_{init}$, a slip velocity moves the contact lines such that at steady state, $\theta\approx\theta_{eq}$. 

\begin{figure}
    \centering
    \includegraphics[width=0.5\linewidth]{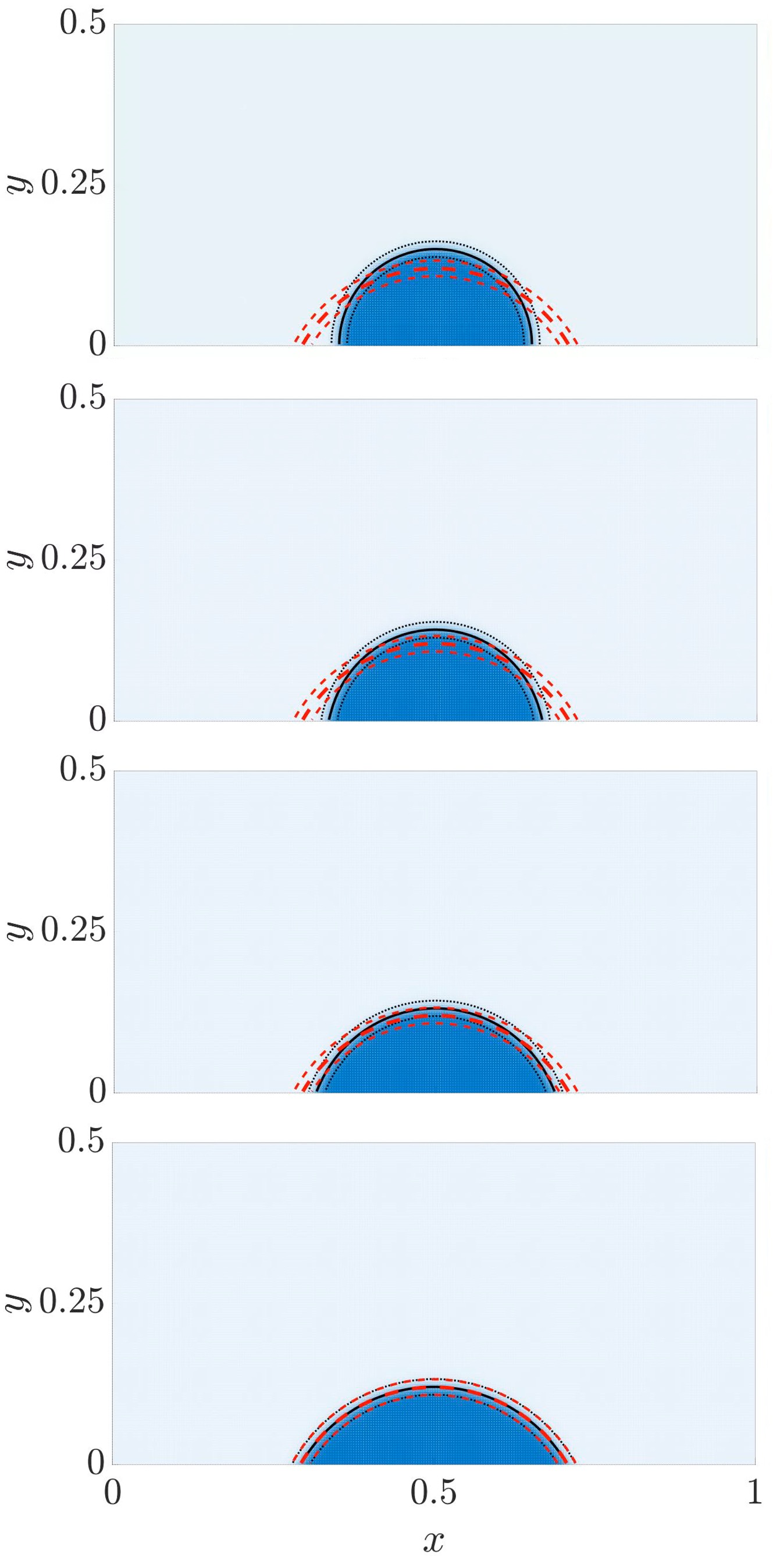}
    \caption{Time series of the evolution of the phase field for the equilibrium drop test case. The black lines mark the $\phi=[0.1,0.5,0.9]$ contours of the numerical solution. The drop is initialized in a zero velocity field as a semi-circle ($\theta_{init}=90^\circ$). The equilibrium contact angle is set to $\theta_{eq}=60^\circ$, and the corresponding $\phi=[0.1,0.5,0.9]$ contours of the exact solution are shown in red. The simulation is run to steady-state, at which point the numerical phase field shows excellent agreement with the exact solution.}
    \label{fig: evolving drop theta_eq=60}
\end{figure}

A time series of the evolution of the phase field is given in Figure \ref{fig: evolving drop theta_eq=60}. The black lines mark the $\phi=[0.1,0.5,0.9]$ contours of the numerical solution, and the dashed red lines mark the $\phi=[0.1,0.5,0.9]$ contours of the exact steady-state solution. These exact contours are from a phase field given again by Equation \ref{eq: prescribed phase field}, but with $\theta=\theta_{eq}$. At steady state, the numerical solution shows excellent agreement with the exact solution. Importantly, all three contour lines match very well with the corresponding contour lines of the exact solution.

\begin{figure}
    \centering
    \includegraphics[width=.7\linewidth]{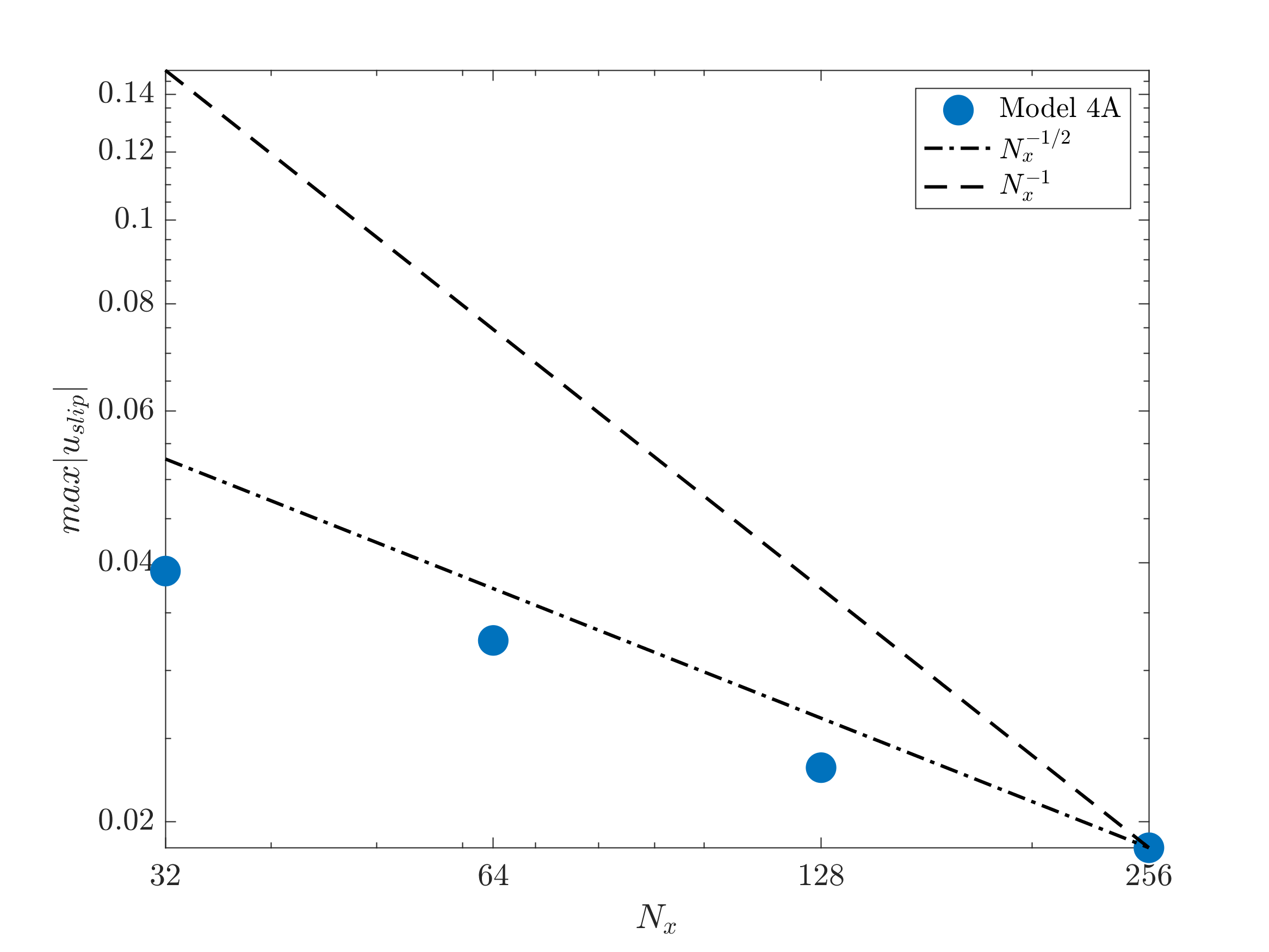}
    \caption{Convergence plot of maximum spurious slip $max(|u_{slip}^Y(x)|)$ as a function of number of points in the $x$ direction $N_x$, for Model 4A. In this convergence, the interface thickness parameter $\epsilon$ is decreased along with the mesh size $\Delta$ according to $\epsilon\sim\Delta^{2/3}$. The dashed black line indicates $N_x^{-1}$ and the dash-dotted black line indicates $N_x^{-1/2}$.}
    \label{fig: max u_slip}
\end{figure}

\begin{figure}
    \centering
    \includegraphics[width=.7\linewidth]{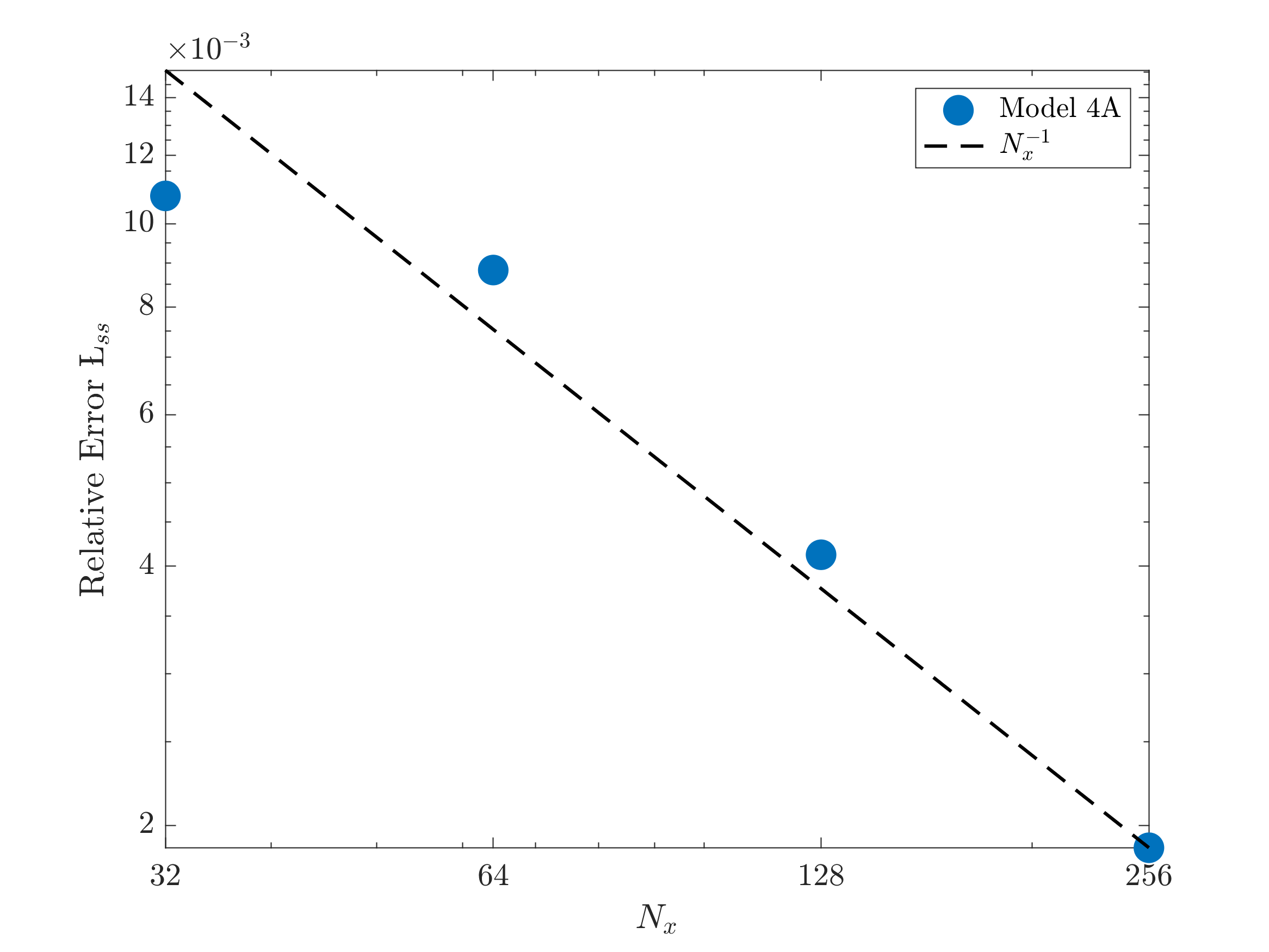}
    \caption{Convergence plot of the relative wetted length error $L_{err}=(L-L_{exact})/L_{exact}$  as a function of number of points in the $x$ direction $N_x$, for Model 4A. In this convergence, the interface thickness parameter $\epsilon$ is decreased along with the mesh size $\Delta$ according to $\epsilon\sim\Delta^{2/3}$. The dashed black line indicates $N_x^{-1}$.}
    \label{fig: relative wetted length error}
\end{figure}

We can also evaluate the convergence of the steady-state results. Similar to the convergence presented in Section \ref{sec: slip boundary condition}, we simultaneously decrease the interface thickness parameter $\epsilon$ and the mesh size $\Delta$, according to $\epsilon\sim\Delta^{2/3}$. The range of $\epsilon$ values spans from $0.7\Delta$ to $1.4\Delta$. The relevant metrics are again the maximum spurious slip $max|u_{slip}|$, as well as the relative wetted length error,
$$
L_{err}=\left|\frac{L-L_{exact}}{L_{exact}}\right|.
$$
$L_{err}$ is a simple metric to evaluate how well the steady-state shape of the drop matches the exact solution. Figure \ref{fig: max u_slip} shows that the maximum spurious slip at steady state is convergent, with an approximate convergence rate of $N_x^{-1/2}$. Figure \ref{fig: relative wetted length error} shows that the relative wetted length error is also convergent, and the rate of convergence is approximately first order $N_x^{-1}$.

\subsection{Two-phase Couette flow}
For this test case, we follow the two-phase Couette flow setup of \cite{qian2005molecular}, who report the parameters in reduced units (in terms of the energy, length, and mass scales of the molecular dynamics simulation). A rectangular drop of length $L_x/2$ and height $L_y$ is initialized in a domain $x\in[0,L_x]$ and $y\in[0,L_y]$, with $L_x=108.8$ and $L_y=13.6$. The initial velocity field is zero. The top and bottom boundaries are walls, on which the contact line treatment is applied and which move at a prescribed wall speed $U_{wall}$. The top wall moves to the right and the bottom wall moves to the left. The left and right boundaries are periodic. The two fluids have identical density $\rho=0.81$ and viscosity $\mu=1.95$, and the surface tension coefficient $\sigma=5.5$. The interface thickness is $\epsilon=0.33\sqrt{2}/2$, which is taken from \cite{qian2005molecular} but adjusted to account for the different definitions of the interface thickness parameter between CH and CDI. The regularization strength is again $\gamma=2.5|\vec{u}|_{max}$, in accordance with \cite{mirjalili2020conservative}.

The initial phase field has interfaces at phase field equilibrium and initial contact angle $\theta_{init}=90^\circ$, and is given by
\begin{equation}
    \phi(x,y) = \frac{1}{2}\left(1 + \tanh\left( \frac{x-L_x/4}{2\epsilon} \right)  \right) + \frac{1}{2}\left(1 + \tanh\left( \frac{3L_x/4 - x}{2\epsilon} \right)  \right) - 1.
    \label{eq: couette flow initial phi}
\end{equation}

\begin{figure}
    \centering
    \includegraphics[width=.7\linewidth]{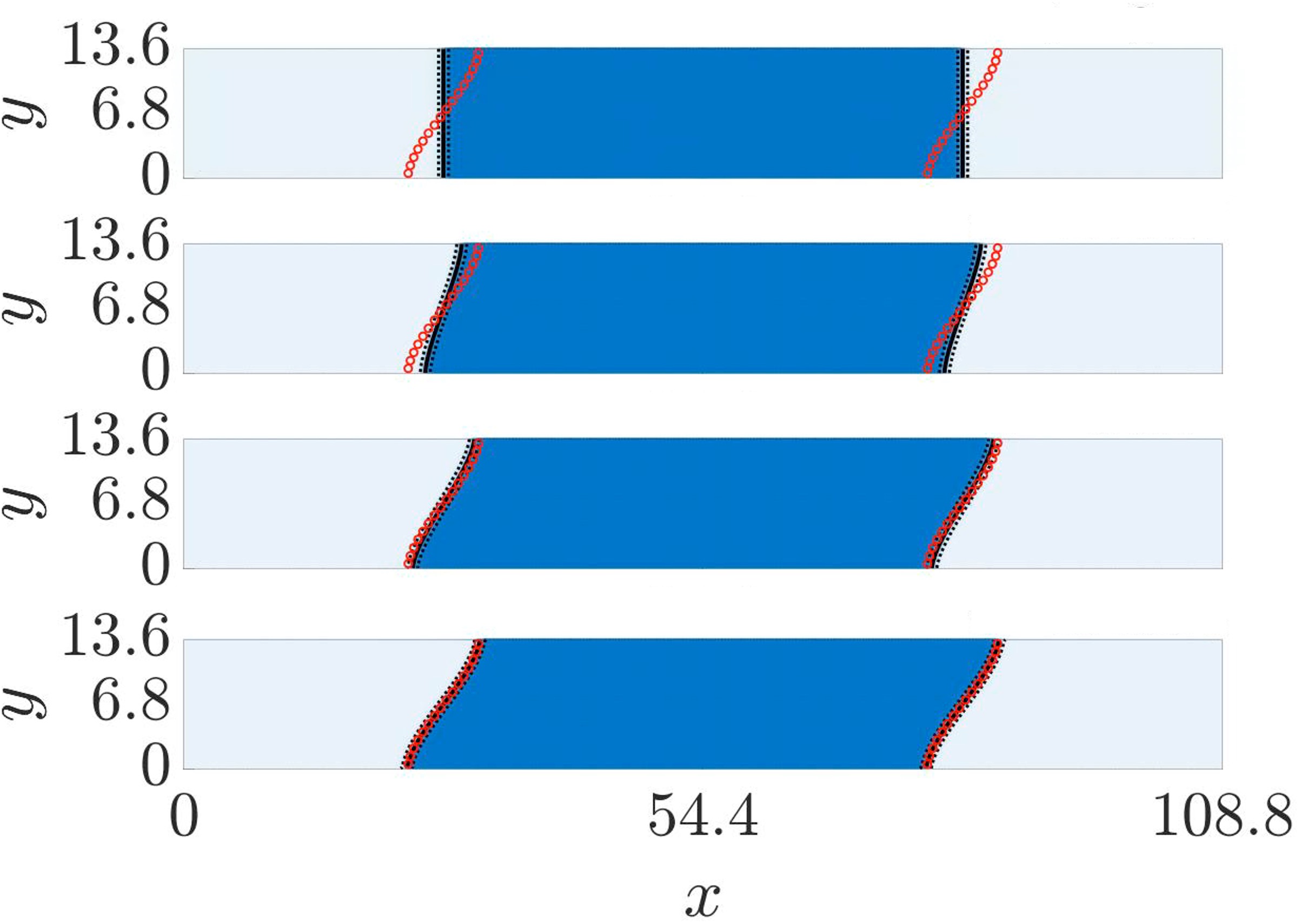}
    \caption{Time series of the evolution of the phase field for the symmetric two-phase Couette flow test case. The black lines mark the $\phi=[0.1,0.5,0.9]$ contours of the numerical solution. The drop is initialized in a zero velocity field with $\theta_{init}=90^\circ$. The equilibrium angle is also set to $\theta_{eq}=90^\circ$, and $\beta_1=\beta_2=1.2$. The top and bottom walls move to the right and left, respectively, at $U_{wall}=0.25$, causing the drop to deform. A steady-state is reached when $\theta$ deviates enough from $\theta_{eq}$ such that the slip velocity counteracts the effect of the wall motion. At steady-state, the numerical solution shows excellent agreement with the molecular dynamics data (red circles) given in \citep{qian2005molecular}.}
    \label{fig: couette flow symmetric}
\end{figure}

\begin{figure}
    \centering
    \includegraphics[width=.7\linewidth]{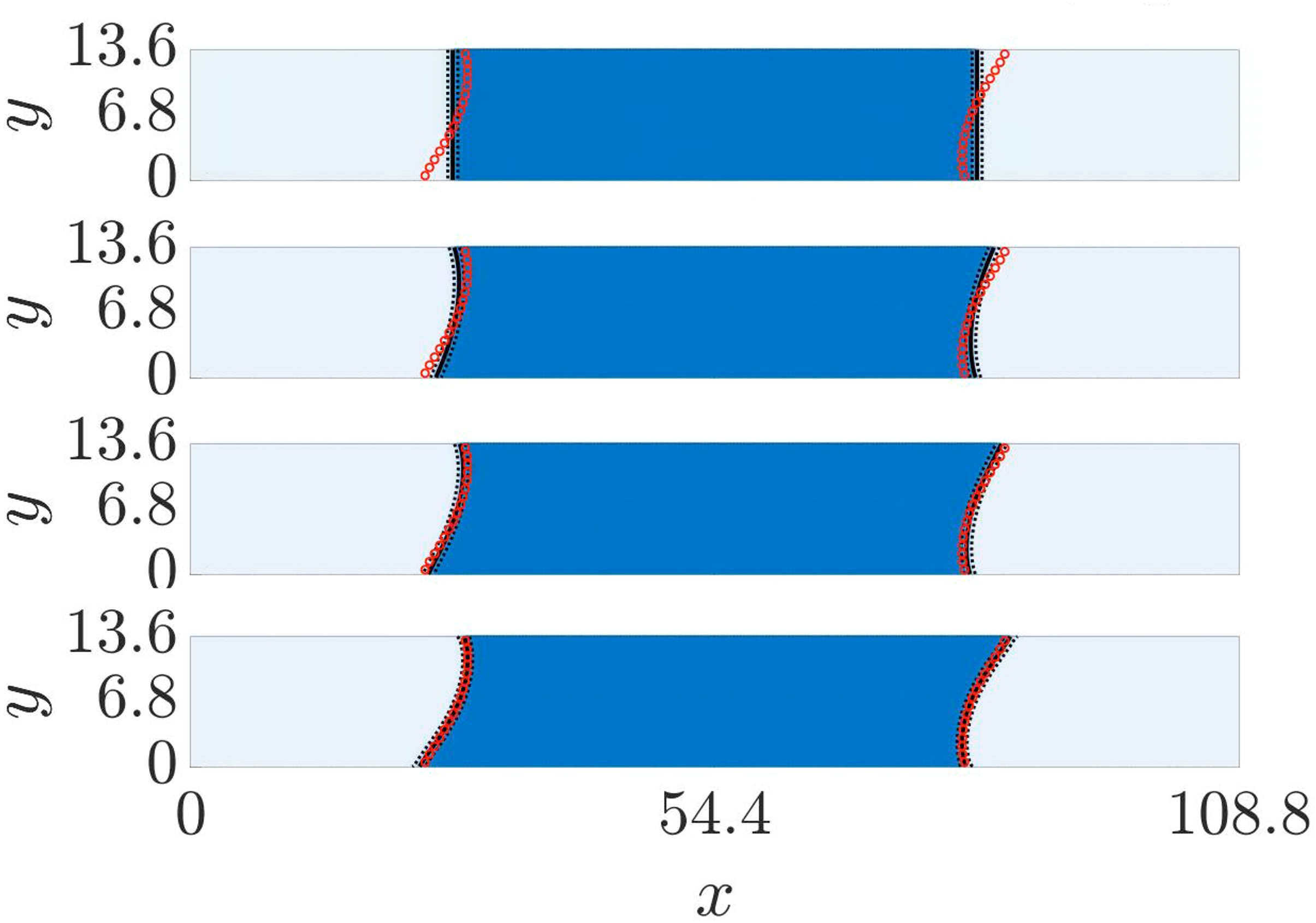}
    \caption{Time series of the evolution of the phase field for the asymmetric two-phase Couette flow test case. The black lines mark the $\phi=[0.1,0.5,0.9]$ contours of the numerical solution. The drop is initialized in a zero velocity field with $\theta_{init}=90^\circ$. The equilibrium angle is set to $\theta_{eq}=\cos^{-1}(0.38)\approx67.7^\circ$, and $\beta_1=1.2$ and $\beta_2=0.591$. The top and bottom walls move to the right and left, respectively, at $U_{wall}=0.20$, causing the drop to deform. A steady-state is reached when $\theta$ deviates enough from $\theta_{eq}$ such that the slip velocity counteracts the effect of the wall motion. At steady-state, the numerical solution shows excellent agreement with the molecular dynamics data (red circles) given in \citep{qian2005molecular}.}
    \label{fig: couette flow asymmetric}
\end{figure}

Results are presented for two variants of this test case. For the ``symmetric" case, the equilibrium contact angle is $\theta_{eq}=90^\circ$, the two fluids have the same wall-fluid friction parameter $\beta_1=\beta_2=1.2$, and the wall speed is $U_{wall}=0.25$. For the ``asymmetric" case, the equilibrium contact angle is $\theta_{eq}=\cos^{-1}(0.38)\approx67.7^\circ$, the two fluids have different wall-fluid friction parameters, $\beta_1=1.2$ and $\beta_2=0.591$, and the wall speed is $U_{wall}=0.20$. A time series of the evolution of the phase field for both cases is given in Figures \ref{fig: couette flow symmetric} and \ref{fig: couette flow asymmetric}. The initially straight interfaces are deformed by the motion of the walls, and for the asymmetric case, also by the fact that the initial contact angles are different from the equilibrium value. In both cases, a steady state is reached when $\theta$ deviates enough from $\theta_{eq}$ such that the slip velocity counteracts the effect of the wall motion. 

At steady state, the numerical results show excellent agreement with the molecular dynamics reference data, shown in red circles \citep{qian2005molecular}. This agreement provides strong validation that our treatment accurately models moving contact line physics.

\section{Conclusion}
\label{sec:conclusions}
This work presents a mass-conserving contact line treatment for second-order conservative phase field methods, based on a no-flux boundary condition for the phase field to ensure mass conservation, and a GNBC-related slip boundary condition for the velocity to model contact line physics. Because the CDI model does not permit a second contact angle boundary condition, ghost values of $\phi$ are never computed or used and near-wall discretization is based on one-sided stencils and extrapolations. In addition, a novel form of the EB surface tension model, along with a related curvature estimate, were found to be discretely advantageous and are employed in this work.

Several versions of the contact line slip $u_{slip}^Y$ were presented and compared in the context of a prescribed equilibrium phase field. It was shown that transforming derivatives of $\phi$ into derivatives of $\psi$ significantly reduces discretization error, and that a curvature correction is necessary to correctly account for the geometry of curved, diffuse interface in phase field equilibrium. Models 4A/B incorporated both modifications and were shown to result in vanishing spurious slip velocities upon convergence.

Proceeding with Model 4A, we presented results for two test cases: the equilibrium drop and two-phase Couette flow. In the former, our steady-state numerical results match the exact diffuse interface solution and show convergence of the spurious slip and the wetted length error. These results provide validation that our treatment correctly models static contact line physics. In the latter, our steady-state numerical results match the molecular dynamics reference data \citep{qian2005molecular}, validating our treatment for moving contact lines as well.


\bibliographystyle{elsarticle-num-names}
\bibliography{main.bib}

\end{document}